\documentclass{ws-ijmpd}
\usepackage[super,compress]{cite}
\begin{document}

\markboth{Zi-Hua Weng}
{Dynamic of astrophysical jets in the complex octonion space}

%
\catchline{}{}{}{}{}
%

\title{Dynamic of astrophysical jets in the complex octonion space
}

\author{Zi-Hua Weng
}

\address{School of Physics and Mechanical \& Electrical Engineering, Xiamen University, Xiamen 361005, China
\\
xmuwzh@xmu.edu.cn}



\maketitle

\begin{history}
\received{Day Month Year}
\revised{Day Month Year}
\end{history}

\begin{abstract}
The paper aims to consider the strength gradient force as the dynamic of astrophysical jets, explaining the movement phenomena of astrophysical jets. J. C. Maxwell applied the quaternion analysis to describe the electromagnetic theory. This encourages others to adopt the complex quaternion and octonion to depict the electromagnetic and gravitational theories. In the complex octonion space, it is capable of deducing the field potential, field strength, field source, angular momentum, torque, force and so forth. As one component of the force, the strength gradient force relates to the gradient of the norm of field strength only, and is independent of not only the direction of field strength but also the mass and electric charge for the test particle. When the strength gradient force is considered as the thrust of the astrophysical jets, one can deduce some movement features of astrophysical jets, including the bipolarity, matter ingredient, precession, symmetric distribution, emitting, collimation, stability, continuing acceleration and so forth. The above results reveal that the strength gradient force is able to be applied to explain the main mechanical features of astrophysical jets, and is the competitive candidate of the dynamic of astrophysical jets.
\end{abstract}


\ccode{PACS numbers:98.38.Fs; 04.50.Kd; 02.10.De; 98.62.Dm; 98.54.Gr; 96.60.Vg.}


\section{Introduction}

The formulaic similarity between the Coulomb force with the Newtonian universal gravitation has been intriguing and puzzling the scholars ever since a long time ago. For many years ago, this puzzle edified some scholars to try to unify all of existing terms of force into one single definition of force, comprehending further the comparability of the Coulomb force and the Newtonian universal gravitation. Finally this research objective was achieved in the electromagnetic and gravitational theories described with the complex octonion in recent years. As one of theoretical applications, the relevant inferences can be applied to explore the dynamic feature of the astrophysical jets.

In his treatise on the electromagnetic field, J. C. Maxwell applied the quaternion analysis and vector terminology to study the electromagnetic field equations. Subsequently some scholars adopted the complex quaternion to depict the physics properties of electromagnetic field equations and of wave equations etc, while other scholars researched the characteristics of gravitational field with the complex quaternion. V. Majernik \cite{majernik} etc adopted the complex quaternion to depict the electromagnetic field theory and the gravitational field theory. W. M. Honig \cite{honig} and A. Singh \cite{singh} applied respectively the complex quaternion to deduce directly the Maxwell's equations in the classical electromagnetic theory. J. Edmonds \cite{edmonds} wrote the wave equation and gravitational theory with the quaternion in the curved space-time. K. Morita \cite{morita} studied the quaternion field theory. S. M. Grusky \cite{grusky} etc researched the time-dependent electromagnetic field by means of the quaternion. H. T. Anastassiu \cite{anastassiu} developed the physics properties of the electromagnetic field with the quaternion. J. G. Winans \cite{winans} discussed the physics quantities with the quaternion. S. Demir \cite{demir} etc applied the hyperbolic quaternion to investigate the electromagnetic theory. F. A. Doria \cite{doria} researched the gravitational theory with the quaternion. A. S. Rawat \cite{rawat} etc harnessed the gravitational field with the quaternion terminology. M. Gogberashvili \cite{gogberashvili} investigated the electromagnetic field with the octonion. V. L. Mironov \cite{mironov} etc made use of the octonion to research the physics properties of electromagnetic equations. O. P. S. Negi \cite{dehnen} etc applied the complex quaternion and octonion to express the Maxwell's equations and the electric current continuity equation.

In the algebra, the ordered couple of complex quaternions compose one complex octonion. On the contrary, one complex octonion space can be separated into two orthogonal subspaces, the complex quaternion space and the $S$-quaternion (short for the second quaternion) space. The results \cite{weng1} reveal that the complex quaternion can be applied to study the feature of gravitational field, while the complex $S$-quaternion can be used to research the feature of electromagnetic field. In a word, the complex quaternion space for the gravitational field is perpendicular to the complex $S$-quaternion space for the electromagnetic field.

In the complex quaternion space for the gravitational field, the gravitational strength consists of two parts, the gravitational acceleration $\textbf{g}$ , and the component $\textbf{b}$ similar to the magnetic flux density. The gravitational acceleration corresponds to the linear acceleration, while the component $\textbf{b}$ corresponds to the precessional angular velocity produced by the gyroscopic torque. The emergence of the linear momentum and the variable gravitational acceleration will induce the component $\textbf{b}$ . The ultrahigh $\textbf{b}$ will bring the remarkable consequence, including the impact on the astrophysical jets. Contrastively, in the classical gravitational theory, there is not the component $\textbf{b}$ . As a result this theory is unable to explain effectively the movement phenomenon of astrophysical jets. Meanwhile the electromagnetic theory described with the complex $S$-quaternion is equivalent to the classical electromagnetic theory, except for the force and the current continuity equation.

In the paper, the complex octonion can be used to depict the physics property of electromagnetic and gravitational fields, including the angular momentum, torque, force and so forth. In the complex octonion space, the definition of force \cite{weng2} comprises the inertial force, Coulomb force, Lorentz force, gravity, energy gradient and so on. As one proportion of the energy gradient, the strength gradient force is defined as the gradient of the norm of electromagnetic strength and of gravitational strength. And this force is distinct from the gradient force in the optical tweezers \cite{yu} or the force in the magnetic mirror \cite{bagryansky}. Furthermore the strength gradient force is independent of the mass or the electric charge for the test particle, and is capable of exerting an influence on any particle of the ordinary matter. Therefore the force can be chosen as the dynamic of astrophysical jets, explaining some movement phenomena relevant to the astrophysical jets.

In the extragalactic system, the jet phenomenon has been observing in a large number of celestial bodies, including the near/far radio galaxies, the quasar radio source \cite{cao}, the Seyfert galaxies and so forth. The active galactic nuclei (AGN) objects are able to emit outward the collimating astrophysical jets. In the structure, the astrophysical jets cover the bipolar-outflow jet, unipolar-outflow jet, precessional jet, reflecting symmetrical jet, mirror-type symmetrical jet and so on. Also the jet phenomenon has been observing in some regions of the Milky Way galaxy, including the Centaurus A, SS433 \cite{cherepashchuk}, Young star, Cygnus X-1 and so on. A great many observations state there are ubiquitous astrophysical jets obviously.

Up to now, the production mechanism of the astrophysical jet is still an unsolved problem \cite{alves} in the radio astronomy. The scholars have been proposing several theoretical models, but these models do not work effectively to explain the movement feature of astrophysical jets. On the basis of the electromagnetic and gravitational theory described with the complex octonion, the paper focuses on proposing one production mechanism of astrophysical jets. In the model of planar circular motion similar to the magnetic dipole moment, the paper explores one new method, considering the strength gradient force as the thrust of the astrophysical jet, in order to explain the dynamic feature of astrophysical jets, including the launching, collimation, stability, procession and so forth.

\section{\label{sec:level1}Octonion field equations}

In the mathematics, the ordered couple of real numbers compose one complex number. Meanwhile the complex number can be separated into two parts, the real number and the imaginary number. Similarly the ordered couple of complex numbers constitute one quaternion. Further the ordered couple of quaternions yield one octonion. On the contrary, one octonion can be broken up into two components, the quaternion and the $S$-quaternion, and even their coordinates may be the complex numbers. In the paper, the quaternion is appropriate to describe the physics property of gravitational field, and the $S$-quaternion is propitious to express the physics property of electromagnetic field.

In the quaternion space $\mathbb{H}_g$ , the basis vector is $\emph{\textbf{i}}_j$ , and the coordinates are $r_0$ and $r_k$ . The radius vector is, $\mathbb{R}_g = i r_0 \emph{\textbf{i}}_0 + \Sigma r_k \emph{\textbf{i}}_k$ , with $\textbf{r} = \Sigma r_k \emph{\textbf{i}}_k$ . The definition of velocity is, $\mathbb{V}_g = v_0 \partial_0 \mathbb{R}_g$ . And it is able to be written as, $\mathbb{V}_g = i v_0 \emph{\textbf{i}}_0 + \Sigma v_k \emph{\textbf{i}}_k$, and $\textbf{v} = \Sigma v_k \emph{\textbf{i}}_k $ , with $v_j$ being the component. The gravitational potential is, $\mathbb{A}_g = i a_0 \emph{\textbf{i}}_0 + \Sigma a_k \emph{\textbf{i}}_k$ , and $\textbf{a} = \Sigma a_k \emph{\textbf{i}}_k $ , with $a_j$ being the component. The definition of gravitational strength is, $\mathbb{F}_g = \square \circ \ \mathbb{A}_g$. And it is written as, $\mathbb{F}_g = f_0 \emph{\textbf{i}}_0 + \Sigma f_k \emph{\textbf{i}}_k$, with $f_j$ being the component. Meanwhile $f_0 = - \partial_0 a_0 + \nabla \cdot \textbf{a}$ , and $\textbf{f} = \Sigma f_k \emph{\textbf{i}}_k$ , with $\textbf{f} = \emph{i} \textbf{g} / v_0 + \textbf{b}$. The gravitational acceleration is, $\textbf{g} / v_0 = \partial_0 \textbf{a} + \nabla a_0$ , and $\textbf{b} = \nabla \times \textbf{a}$ . The definition of gravitational source is, $\mu_g \mathbb{S}_g = - \square^* \circ \mathbb{F}_g$ . And it can be expressed as, $\mathbb{S}_g = i s_0 \emph{\textbf{i}}_0 + \Sigma s_k \emph{\textbf{i}}_k$, and $\textbf{s} = \Sigma s_k \emph{\textbf{i}}_k$ , with $s_j$ being the component. Herein $v_0 = \partial r_0 / \partial t$ , $t$ is the time, and $v_0$ is the speed of light, $c$. The gravitational constant is, $\mu_g < 0$ . $i$ is the imaginary unit. $r_j, ~v_j, ~a_j, ~s_j$, and $f_0$ are all real. $f_k$ is the complex number. The quaternion operator is, $\square = \emph{i} \emph{\textbf{i}}_0 \partial_0 + \Sigma \emph{\textbf{i}}_k \partial_k$ , and $ \nabla = \Sigma \emph{\textbf{i}}_k \partial_k$, with $ \partial_j = \partial / \partial r_j$ . $\circ$ is the octonion multiplication, and $*$ is the octonion conjugate. $\emph{\textbf{i}}_0 = 1$. $\emph{\textbf{i}}_k^2 = -1$. $j = 0, 1, 2, 3$. $k = 1, 2, 3$.

Choosing the gauge condition, $- f_0 = 0$, one can substitute $\mathbb{S}_g$ and $\mathbb{F}_g$ into the definition of gravitational source, $\mu_g \mathbb{S}_g = - \square^* \circ \mathbb{F}_g$ . In the left side of equal sign, the coordinate of scalar part (relevant to $\emph{\textbf{i}}_0$ ) comprises only the imaginary part, while that of vector part (relevant to $\emph{\textbf{i}}_k$ ) contains only the real part. In the right side of equal sign, the coordinate of scalar part comprises the real and imaginary parts, while that of vector part contains the real and imaginary parts also. Comparing both sides of the equal sign yields four equations, which compose the gravitational field equations (Table I). In the case for one single particle, a comparison with the classical gravitational theory reveals that, $\mathbb{S}_g = m \mathbb{V}_g$ , with $m$ being the mass density. When $\textbf{b} = 0$ and $\textbf{a} = 0$, one of the gravitational field equations will be able to be degenerated into the Newton's law of universal gravitation in the classical gravitational theory.

In the $S$-quaternion space $\mathbb{H}_e$ , the basis vector is $\emph{\textbf{I}}_j$ , and the coordinates are $R_0$ and $R_k$ . The radius vector is, $\mathbb{R}_e = i R_0 \emph{\textbf{I}}_0 + \Sigma R_k \emph{\textbf{I}}_k$, with $\textbf{R} = \Sigma R_k \emph{\textbf{I}}_k$, and $\textbf{R}_0 = R_0 \emph{\textbf{I}}_0$. The definition of velocity is, $\mathbb{V}_e = v_0 \partial_0 \mathbb{R}_e$ . And it is able to be written as, $\mathbb{V}_e = i V_0 \emph{\textbf{I}}_0 + \Sigma V_k \emph{\textbf{I}}_k $, and $\textbf{V} = \Sigma V_k \emph{\textbf{I}}_k$, and $\textbf{V}_0 = V_0 \emph{\textbf{I}}_0$ , with $V_j$ being the component. The electromagnetic potential is, $\mathbb{A}_e = i A_0 \emph{\textbf{I}}_0 + \Sigma A_k \emph{\textbf{I}}_k $, and $\textbf{A} = \Sigma A_k \emph{\textbf{I}}_k $, and $\textbf{A}_0 = A_0 \emph{\textbf{I}}_0$ , with $A_j$ being the component. The definition of electromagnetic strength is, $\mathbb{F}_e = \square \circ \mathbb{A}_e$ . And it can be written as, $\mathbb{F}_e = F_0 \emph{\textbf{I}}_0 + \Sigma F_k \emph{\textbf{I}}_k$ , with $F_j$ being the component. Meanwhile $\textbf{F}_0 = - \partial_0 \textbf{A}_0 + \nabla \cdot \textbf{A}$ , and $\textbf{F}_0 = F_0 \emph{\textbf{I}}_0 $, and $\textbf{F} = \Sigma F_k \emph{\textbf{I}}_k$, with $\textbf{F} = \emph{i} \textbf{E} / v_0 + \textbf{B}$ . The electric field intensity is, $\textbf{E} / v_0 = \partial_0 \textbf{A} + \nabla \circ \textbf{A}_0$, and the magnetic flux density is, $\textbf{B} = \nabla \times \textbf{A}$ . The definition of electromagnetic source is, $\mu_e \mathbb{S}_e = - \square^* \circ \mathbb{F}_e$ . And it can be expressed as, $\mathbb{S}_e = i S_0 \emph{\textbf{I}}_0 + \Sigma S_k \emph{\textbf{I}}_k$, and $\textbf{S} = \Sigma S_k \emph{\textbf{I}}_k$, and $\textbf{S}_0 = S_0 \emph{\textbf{I}}_0$ , with $S_j$ being the component. Herein $\textbf{V}_0 = \partial \textbf{R}_0 / \partial t$ . The electromagnetic constant is, $\mu_e > 0$ . $R_j, ~V_j, ~A_j, ~S_j$, and $F_0$ are all real. $F_k$ is the complex number. $\emph{\textbf{I}}_j = \emph{\textbf{i}}_j \circ \emph{\textbf{I}}_0$. $\emph{\textbf{I}}_j^2 = -1$.

Similarly choosing the gauge condition, $- \textbf{F}_0 = 0$, one can substitute $\mathbb{S}_e$ and $\mathbb{F}_e$ into the definition of electromagnetic source, $\mu_e \mathbb{S}_e = - \square^* \circ \mathbb{F}_e$ . In the left side of equal sign, the coordinate of scalar-like part (relevant to $\emph{\textbf{I}}_0$ ) comprises only the imaginary part, while that of vector-like part (relevant to $\emph{\textbf{I}}_k$ ) contains only the real part. In the right side of equal sign, the coordinate of scalar-like part comprises the real and imaginary parts, while that of vector-like part contains the real and imaginary parts also. Comparing both sides of the equal sign will reason out four equations, which compose the electromagnetic field equations. In the case for one single charged particle, a comparison with the classical electromagnetic theory reveals that, $\mathbb{S}_e = \rho \mathbb{V}_e$ , with $\rho$ being the density of electric charge. The above electromagnetic field equations are identical with the Maxwell's equations in the classical electromagnetic theory.

In the octonion space $\mathbb{O}$, the octonion radius vector is, $\mathbb{R} = \mathbb{R}_g + k_{eg} \mathbb{R}_e$ , and the octonion velocity is, $\mathbb{V} = \mathbb{V}_g + k_{eg} \mathbb{V}_e$ . The octonion field potential is, $\mathbb{A} = \mathbb{A}_g + k_{eg} \mathbb{A}_e$, and the octonion field strength is, $\mathbb{F} = \mathbb{F}_g + k_{eg} \mathbb{F}_e$ . The definition of octonion field source is,
\begin{eqnarray}
\mu \mathbb{S} && = - ( \emph{i} \mathbb{F} / v_0 + \square )^* \circ  \mathbb{F}
\nonumber \\
&& = \mu_g \mathbb{S}_g + k_{eg} \mu_e \mathbb{S}_e - ( \emph{i} \mathbb{F} / v_0 )^* \circ \mathbb{F} ~,
\end{eqnarray}
where $\mu$ and $k_{eg}$ are the coefficients. According to the coefficient $k_{eg}$ and the basis vector $( \emph{\textbf{i}}_j , \emph{\textbf{I}}_j )$, the above can be separated into two parts, the definition of gravitational source, and the definition of electromagnetic source.

On the analogy of the definition of complex coordinate system, one can define the coordinate of octonion, which involves the quaternion and $S$-quaternion simultaneously. In the octonion coordinate system, the octonion physics quantity can be defined as $ \{ ( \emph{i} c_0 + \emph{i} d_0 \textbf{\emph{I}}_0 ) \circ \emph{\textbf{i}}_0 + \Sigma ( c_k + d_k \textbf{\emph{I}}_0^* ) \circ \emph{\textbf{i}}_k \} $. It means that there are the quaternion coordinate $c_k$ and the $S$-quaternion coordinate $d_k \emph{\textbf{I}}_0^*$ for the basis vector $\emph{\textbf{i}}_k$, while the quaternion coordinate $c_0$ and the $S$-quaternion coordinate $d_0 \emph{\textbf{I}}_0$ for the basis vector $\emph{\textbf{i}}_0$. Herein $c_j$ and $d_j$ are all real.

\begin{table}[ph]
\tbl{The multiplication of the quaternion operator and octonion physics quantity.}
{\begin{tabular}{@{}ll@{}}
\hline
definition                  &   expression~meaning                                   \\
\hline
$\nabla \cdot \textbf{a}$   &  $-(\partial_1 a_1 + \partial_2 a_2 + \partial_3 a_3)$ \\
$\nabla \times \textbf{a}$  &  $\emph{\textbf{i}}_1 ( \partial_2 a_3
                                 - \partial_3 a_2 ) + \emph{\textbf{i}}_2 ( \partial_3 a_1
                                 - \partial_1 a_3 )
                                 + \emph{\textbf{i}}_3 ( \partial_1 a_2
                                 - \partial_2 a_1 )$                                 \\
$\nabla a_0$                &  $\emph{\textbf{i}}_1 \partial_1 a_0
                                 + \emph{\textbf{i}}_2 \partial_2 a_0
                                 + \emph{\textbf{i}}_3 \partial_3 a_0  $             \\
$\partial_0 \textbf{a}$     &  $\emph{\textbf{i}}_1 \partial_0 a_1
                                 + \emph{\textbf{i}}_2 \partial_0 a_2
                                 + \emph{\textbf{i}}_3 \partial_0 a_3  $             \\

$\nabla \cdot \textbf{A}$   &  $-(\partial_1 A_1 + \partial_2 A_2 + \partial_3 A_3) \emph{\textbf{I}}_0 $  \\
$\nabla \times \textbf{A}$  &  $-\emph{\textbf{I}}_1 ( \partial_2
                                 A_3 - \partial_3 A_2 ) - \emph{\textbf{I}}_2 ( \partial_3 A_1
                                 - \partial_1 A_3 )
                                 - \emph{\textbf{I}}_3 ( \partial_1 A_2
                                 - \partial_2 A_1 )$                                 \\
$\nabla \circ \textbf{A}_0$ &  $\emph{\textbf{I}}_1 \partial_1 A_0
                                 + \emph{\textbf{I}}_2 \partial_2 A_0
                                 + \emph{\textbf{I}}_3 \partial_3 A_0  $             \\
$\partial_0 \textbf{A}$     &  $\emph{\textbf{I}}_1 \partial_0 A_1
                                 + \emph{\textbf{I}}_2 \partial_0 A_2
                                 + \emph{\textbf{I}}_3 \partial_0 A_3  $             \\
\hline
\end{tabular}}
\end{table}

\section{\label{sec:level1}Octonion angular momentum}

From the octonion field source, the octonion linear momentum density $\mathbb{P}$ can be defined as,
\begin{equation}
\mathbb{P} = \mu \mathbb{S} / \mu_g ~,
\end{equation}
where $\mathbb{P} = \mathbb{P}_g + k_{eg} \mathbb{P}_e$ . In the quaternion space $\mathbb{H}_g$, the component of linear momentum density is, $\mathbb{P}_g = \{ \mu_g \mathbb{S}_g - ( \emph{i} \mathbb{F} / v_0 )^* \circ \mathbb{F} \} / \mu_g$ . $\mathbb{P}_g = i p_0 + \textbf{p}$ , and $\textbf{p} = \Sigma p_k \emph{\textbf{i}}_k$. In the $S$-quaternion space $\mathbb{H}_e$ , the component of linear momentum density is, $\mathbb{P}_e = \mu_e \mathbb{S}_e / \mu_g$ . $\mathbb{P}_e = i \textbf{P}_0 + \textbf{P}$ , with $\textbf{P} = \Sigma P_k \emph{\textbf{I}}_k$ , and $\textbf{P}_0 = P_0 \emph{\textbf{I}}_0$ . $p_j$ and $P_j$ are all real.

From the octonion linear momentum, the octonion angular momentum density $\mathbb{L}$ is defined as,
\begin{eqnarray}
\mathbb{L} = (\mathbb{R} + k_{rx} \mathbb{X})^\times \circ \mathbb{P}~,
\end{eqnarray}
where $k_{rx}$ is the coefficient. $\times$ is the complex conjugate. $\mathbb{L} = \mathbb{L}_g + k_{eg} \mathbb{L}_e$ . $\mathbb{X}$ is the integrating function of the field potential $\mathbb{A}$ . That is, $\mathbb{A} = i \square^\times \circ \mathbb{X}$ . $\mathbb{X} = \mathbb{X}_g + k_{eg} \mathbb{X}_e$. $\mathbb{X}_g = \emph{i} x_0 + \textbf{x}$ , and $\textbf{x} = \Sigma x_k \emph{\textbf{i}}_k$ . $\mathbb{X}_e = \emph{i} \textbf{X}_0 + \textbf{X}$ , with $\textbf{X}_0 = X_0 \emph{\textbf{I}}_0$ , and $\textbf{X} = \Sigma X_k \emph{\textbf{I}}_k$. $x_j$ and $X_j$ are all real.

In the quaternion space $\mathbb{H}_g$ , the component of angular momentum density is,
\begin{eqnarray}
\mathbb{L}_g = (\mathbb{R}_g^+)^\times \circ \mathbb{P}_g + k_{eg}^2 (\mathbb{R}_e^+)^\times \circ \mathbb{P}_e~,
\end{eqnarray}
where $\mathbb{L}_g = L_{10} + \emph{i} \textbf{L}_1^i + \textbf{L}_1$. $\textbf{L}_1 = \textbf{r}^+ \times \textbf{p} + k_{eg}^2 \textbf{R}^+ \times \textbf{P} $, covers the conventional angular momentum density etc in the classical mechanics. $\textbf{L}_1^i = \Sigma L_{1k}^i \emph{\textbf{i}}_k$. $\textbf{L}_1 = \Sigma L_{1k} \emph{\textbf{i}}_k$. $L_{1j}$ and $L_{1k}^i$ are all real. $\mathbb{R}^+ = \mathbb{R} + k_{rx} \mathbb{X}$, and $\mathbb{R}^+ = \mathbb{R}_g^+ + k_{eg} \mathbb{R}_e^+$. $\mathbb{R}_g^+ = \emph{i} r_0^+ + \textbf{r}^+ $, and $\mathbb{R}_e^+ = \emph{i} \textbf{R}_0^+ + \textbf{R}^+$. $r_j^+ = r_j + k_{rx} x_j$. $R_j^+ = R_j + k_{rx} X_j$. $\textbf{r}^+ = \Sigma r_k^+ \emph{\textbf{i}}_k$. $\textbf{R}_0^+ = R_0^+ \emph{\textbf{I}}_0$, $\textbf{R}^+ = \Sigma R_k^+ \emph{\textbf{I}}_k$.

In the $S$-quaternion space $\mathbb{H}_e$ , the component of angular momentum density is,
\begin{eqnarray}
\mathbb{L}_e =  (\mathbb{R}_g^+)^\times \circ \mathbb{P}_e + (\mathbb{R}_e^+)^\times \circ \mathbb{P}_g ~,
\end{eqnarray}
where $\mathbb{L}_e = \textbf{L}_{20} + \emph{i} \textbf{L}_2^i + \textbf{L}_2$ . $\textbf{L}_2 = \textbf{r}^+ \times \textbf{P} + \textbf{R}^+ \times \textbf{p}$ , covers the magnetic dipole moment etc. $\textbf{L}_{20} = L_{20} \emph{\textbf{I}}_0$ . $\textbf{L}_2^i = \Sigma L_{2k}^i \emph{\textbf{I}}_k$ , includes the electric dipole moment etc. $\textbf{L}_2 = \Sigma L_{2k} \emph{\textbf{I}}_k$ . $L_{2j}$ and $L_{2k}^i$ are all real.

Being similar to other components (the angular momentum and the electric/magnetic dipole moment) of the octonion angular momentum, one component, $(r_0 \mathbb{P})$, will play an important role in the following discussion, expressing the force formula and so forth in the classical mechanics.

\section{\label{sec:level1}Octonion torque}

The octonion torque density $\mathbb{W}$ can be define from the octonion angular momentum,
\begin{equation}
\mathbb{W} = - v_0 ( \emph{i} \mathbb{F} / v_0 + \square ) \circ \mathbb{L} ~,
\end{equation}
where $\mathbb{W} = \mathbb{W}_g + k_{eg} \mathbb{W}_e$ .

In the quaternion space $\mathbb{H}_g$ , the component of torque density is,
\begin{equation}
\mathbb{W}_g = - ( \emph{i} \mathbb{F}_g \circ \mathbb{L}_g + \emph{i} k_{eg}^2 \mathbb{F}_e \circ \mathbb{L}_e + v_0 \square \circ \mathbb{L}_g )~,
\end{equation}
where $\mathbb{W}_g = \emph{i} W_{10}^i + W_{10} + \emph{i} \textbf{W}_1^i + \textbf{W}_1$ . $W_{10}^i$ is the energy density. The energy includes the kinetic energy, potential energy, work, and the interacting energy between the particles with the field, and so on. $-\textbf{W}_1^i$ is the torque density. The torque covers the torque produced by the applied force and so forth. $\textbf{W}_1$ is the curl of angular momentum density, while $W_{10}$ is the divergence of angular momentum density. $\textbf{W}_1 = \Sigma W_{1k} \emph{\textbf{i}}_k$, $\textbf{W}_1^i = \Sigma W_{1k}^i \emph{\textbf{i}}_k$. $W_{1j}$ and $W_{1j}^i$ are all real.

In the energy density $W_{10}^i$ , it is associated with the definition of field potential. The gravitational potential is defined as, $\mathbb{A}_g = \emph{i} \square^\times \circ \mathbb{X}_g$ . $\mathbb{A}_g = \emph{i} a_0 + \textbf{a}$, with $a_0 = \partial_0 x_0 + \nabla \cdot \textbf{x}$, and $\textbf{a} = \partial_0 \textbf{x} - \nabla x_0$ . While the electromagnetic potential is defined as, $\mathbb{A}_e = \emph{i} \square^\times \circ \mathbb{X}_e$ . $\mathbb{A}_e = \emph{i} \textbf{A}_0 + \textbf{A}$, with $\textbf{A}_0 = \partial_0 \textbf{X}_0 + \nabla \cdot \textbf{X}$, and $\textbf{A} = \partial_0 \textbf{X} - \nabla \circ \textbf{X}_0$. The gauge conditions are chosen as, $\nabla \times \textbf{x} = 0$ and $\nabla \times \textbf{X} = 0$. When the radius vector, $\textbf{r}$, is three-dimensional, the term $( W_{10}^i / 2 )$ is the conventional energy density in the classical field theory. And it is stated, $k_{rx} = 1 / v_0$, to compare with the potential energy in the classical gravitational and electromagnetic fields.

In the $S$-quaternion space $\mathbb{H}_e$ , the component of torque density is,
\begin{equation}
\mathbb{W}_e = - ( \emph{i} \mathbb{F}_g \circ \mathbb{L}_e + \emph{i} \mathbb{F}_e \circ \mathbb{L}_g + v_0 \square \circ \mathbb{L}_e ) ~,
\end{equation}
where $\mathbb{W}_e = \emph{i} \textbf{W}_{20}^i + \textbf{W}_{20} + \emph{i} \textbf{W}_2^i + \textbf{W}_2$ . $\textbf{W}_{20}$ covers the divergence of magnetic dipole moment, while $\textbf{W}_2$ includes the curl of magnetic dipole moment and the derivative of electric dipole moment. $\textbf{W}_{20} = W_{20} \emph{\textbf{I}}_0$, $\textbf{W}_{20}^i = W_{20}^i \emph{\textbf{I}}_0$. $\textbf{W}_2 = \Sigma W_{2k} \emph{\textbf{I}}_k$, $\textbf{W}_2^i = \Sigma W_{2k}^i \emph{\textbf{I}}_k$. $W_{2j}$ and $W_{2j}^i$ are all real.

The octonion torque component $(v_0 \mathbb{P})$ comes from the contribution of the octonion angular momentum component $(r_0 \mathbb{P})$. On the analogy of the some torque components (the torque and the energy), the component $(v_0 \mathbb{P})$ has an indispensable influence on the expression of the force equilibrium equation and the precessional angular velocity and so forth (Table II).

\section{\label{sec:level1}Octonion Force}

From the octonion torque, the octonion force density $\mathbb{N}$ is defined as,
\begin{equation}
\mathbb{N} = - ( \emph{i} \mathbb{F} / v_0 + \square ) \circ \mathbb{W}~,
\end{equation}
where $\mathbb{N} = \mathbb{N}_g + k_{eg} \mathbb{N}_e$ . In the quaternion space $\mathbb{H}_g$ , the component of force density is, $\mathbb{N}_g = - ( \emph{i} \mathbb{F}_g \circ \mathbb{W}_g / v_0 + \emph{i} k_{eg}^2 \mathbb{F}_e \circ \mathbb{W}_e / v_0 + \square \circ \mathbb{W}_g ) $, which can infer the force, precessional angular velocity, mass continuity equation and so on. In the $S$-quaternion space $\mathbb{H}_e$ , the component of force density is, $\mathbb{N}_e = - ( \emph{i} \mathbb{F}_g \circ \mathbb{W}_e / v_0 + \emph{i} \mathbb{F}_e \circ \mathbb{W}_g / v_0 + \square \circ \mathbb{W}_e ) $, which is capable of deducing the current continuity equation and so forth.

\subsection{\label{sec:level2}Force equilibrium equation}

In the quaternion space $\mathbb{H}_g$ , the component $\mathbb{N}_g$ of force density can be expanded into
\begin{equation}
\mathbb{N}_g = \emph{i} N_{10}^i + N_{10} + \emph{i} \textbf{N}_1^i + \textbf{N}_1 ~,
\end{equation}
where $N_{10}^i$ is the torque divergence. $N_{10}$ is the power density. $\textbf{N}_1^i$ is the force density. $\textbf{N}_1$ is the torque derivative. $\textbf{N}_1 = \Sigma N_{1k} \emph{\textbf{i}}_k$, $\textbf{N}_1^i = \Sigma N_{1k}^i \emph{\textbf{i}}_k$. $N_{1j}$ and $N_{1j}^i$ are all real.

When $\mathbb{N}_g = 0$, the mass continuity equation can be derived from $N_{10} = 0$. The force equilibrium equation is able to be deduced from $\textbf{N}_1^i = 0$. Further it can be degenerated into the Newton Second Law. From $\textbf{N}_1 = 0$, one can infer the velocity curl of the test particle, including the angular velocity of Larmor precession for one charged particle.

From the force equilibrium equation, $\textbf{N}_1^i = 0$, it is able to obtain approximately,
\begin{eqnarray}
&& - \partial_0 (\textbf{p} v_0) + k_{eg}^2 ( \frac{ \textbf{E} \circ \textbf{P}_0 } { v_0 }\, - \textbf{B} \times \textbf{P} )
+  \frac{  L_{10} ( \textbf{g} \times \textbf{b} + k_{eg}^2 \textbf{E} \times \textbf{B} ) } { v_0^2 k_p }\,
\nonumber \\
&&  - \textbf{b} \times \textbf{p} + \{ \frac{ p_0 } { v_0 }\, + \frac{ W_E } { k_p v_0^2 }\, \} \textbf{g}  - \nabla ( p_0 v_0 +  \frac{ W_E } { k_p }\, ) = 0 ~,
\end{eqnarray}
where $\{ p_0 + W_E / (k_p v_0)\} \textbf{g} / v_0$ is the gravity density, $\partial_0 (- \textbf{p} v_0)$ is the inertial force density, $k_{eg}^2 ( \textbf{E} \circ \textbf{P}_0 / v_0 - \textbf{B} \times \textbf{P} )$ is the density of electromagnetic force. $- \nabla (p_0 v_0 + W_E / k_p )$ is the gradient of energy density. $L_{10} ( k_{eg}^2 \textbf{E} \times \textbf{B} ) / ( v_0^2 k_p )$ is in direct proportion to the electromagnetic momentum. The sum of the gravitational potential energy and electromagnetic potential energy is, $W_E = - \{ ( p_0 a_0 + \textbf{a} \cdot \textbf{p}) + k_{eg}^2 ( \textbf{A}_0 \circ \textbf{P}_0 + \textbf{A} \cdot \textbf{P} ) \}$. $k_p = (k - 1)$ is the coefficient, with $k$ being the dimension of vector $\textbf{r}$ . Comparing with the electromagnetic force in the classical field theory states that, $k_{eg}^2 = \mu_g / \mu_e < 0$.

\subsection{\label{sec:level2}Strength gradient force}

As discussed in the previous section, the energy gradient is one component of the force. Sometimes the energy gradient is called as the gradient force also, and that it contains a few terms obviously. In Eq.(11) the gradient force is,
\begin{eqnarray}
\textbf{N}_B = - \nabla ( p_0 v_0 +  \frac{ W_E } { k_p }\, ) ~.
\end{eqnarray}

When $\nabla m = 0$ and $k = 3$, the above is reduced to
\begin{eqnarray}
\textbf{N}_B = && \nabla ( \frac{ b^2 } { \mu_g }\, - \frac{ g^2 } { v_0^2 \mu_g }\, ) - \nabla ( \frac{ b^2 } { 2 \mu_g }\, - \frac{ g^2 } { 2 v_0^2 \mu_g }\, )
\nonumber
\\
&&
+ \nabla ( \frac{ B^2 } { \mu_e  }\, - \frac{ E^2 } { v_0^2 \mu_e } ) - \nabla ( \frac{ B^2 } { 2 \mu_e  }\, + \frac{ E^2 } { 2 v_0^2 \mu_e } ) ~ ,
\end{eqnarray}
where $W_E = ( b^2 - g^2 / v_0^2 ) / (2 \mu_g) + ( B^2 + E^2 / v_0^2 ) / (2 \mu_e)$. $m^\prime = - \mathbb{F}^* \circ \mathbb{F} / ( v_0^2 \mu_g ) = - ( b^2 - g^2 / v_0^2 ) / ( v_0^2 \mu_g ) - ( B^2 - E^2 / v_0^2 ) / ( v_0^2 \mu_e )$. $b^2 = \textbf{b}^* \cdot \textbf{b}$ , $g^2 = \textbf{g}^* \cdot \textbf{g}$ . $E^2 = \textbf{E}^* \cdot \textbf{E}$, $B^2 = \textbf{B}^* \cdot \textbf{B}$ . In order to distinguish from other terms of the gradient force, $\textbf{N}_B$ is called as the strength gradient force for the moment in the paper.

When there is only the gravitational strength component $\textbf{b}$ , the above will be reduced to
\begin{eqnarray}
\textbf{N}_B = \nabla \frac{ b^2 } { 2 \mu_g }\, ~.
\end{eqnarray}

In the octonion space, from the definition of field strength and of field equations, it is able to deduce the d'Alembert equation for the gravitational and electromagnetic fields. It claims that the gravitational potential is determined by the mass and linear momentum, while the electromagnetic potential is dependent on the electric charge and current. Moreover the gravitational strength component $\textbf{b}$ is decided by the linear momentum and the fluctuant $\textbf{g}$ .

On one orbital plane, one particle is in an uniform circular motion, and the particle movement produces the gravitational strength component $\textbf{b}$ . By analogy with the magnetic dipole moment, the distribution of component $\textbf{b}$ is symmetrical with respect to the symmetric axis A-A. And that the symmetric axis A-A goes through the center point of circular orbit, and is perpendicular to the orbital plane.

According to the above, the magnitude of strength gradient force $\textbf{N}_B$ is in direct proportion to the magnitude $b$ as well as $\nabla b$ . However $\nabla b^2$ is independent to the direction of gravitational strength component $\textbf{b}$ . It means that the strength gradient force $\textbf{N}_B$ will push particles of the ordinary matter to move outward, along the two ends of the symmetric axis A-A simultaneously.

The movement model of particle is complicated comparatively, when $ \partial m \neq 0$. The distribution of gravitational strength component $\textbf{b}$ , which is produced by the particle movement, will be intricate. Further it will result in the launching mode of particles, under the influence of the strength gradient force $\textbf{N}_B$ , to be more complex than the above discussion.

\begin{table}[h]
\tbl{Some definitions of the physics quantity in the gravitational and electromagnetic fields described with the complex octonion.}
{\begin{tabular}{@{}ll@{}}
\hline
octonion~physics~quantity ~~~~~~~~    &   definition                                                                                 \\
\hline
quaternion~operator                   &  $\square = i \partial_0 + \Sigma \emph{\textbf{i}}_k \partial_k$                            \\
radius~vector                         &  $\mathbb{R} = \mathbb{R}_g + k_{eg} \mathbb{R}_e  $                                         \\
integrating~function                  &  $\mathbb{X} = \mathbb{X}_g + k_{eg} \mathbb{X}_e  $                                         \\
field~potential                       &  $\mathbb{A} = i \square^\times \circ \mathbb{X}  $                                          \\
field~strength                        &  $\mathbb{F} = \square \circ \mathbb{A}  $                                                   \\
field~source                          &  $\mu \mathbb{S} = - ( i \mathbb{F} / v_0 + \square )^* \circ \mathbb{F} $                   \\

linear~momentum                       &  $\mathbb{P} = \mu \mathbb{S} / \mu_g $                                                      \\
angular~momentum                      &  $\mathbb{L} = ( \mathbb{R} + k_{rx} \mathbb{X} )^\times \circ \mathbb{P} $                  \\
octonion~torque                       &  $\mathbb{W} = - v_0 ( i \mathbb{F} / v_0 + \square ) \circ \mathbb{L} $          ~~         \\
octonion~force                        &  $\mathbb{N} = - ( i \mathbb{F} / v_0 + \square ) \circ \mathbb{W} $                         \\
\hline
\end{tabular}}
\end{table}

\section{\label{sec:level1}Astrophysical jets}

In the planar circular motion similar to the magnetic dipole moment, the strength gradient force, which is relevant to the distribution of gravitational strength component $\textbf{b}$ , is able to be applied to explore the phenomena of astrophysical jets, especially the astrophysical jet emitted along the precessional axis (or symmetric axis) A-A, near the rotational axis of the celestial body $M$.

\subsection{\label{sec:level2}Jet launching}

In the vacuum, there are one large-mass celestial body $M$, and one small-mass celestial body $m$ . Under the attraction of gravity, the celestial body $m$ moves around the celestial body $M$ in the elliptical orbit. The rotation of celestial body $M$ and the revolution of celestial body $m$ produce jointly the gravitational strength component $\textbf{b}$ . The distribution of component $\textbf{b}$ yields the strength gradient force $\textbf{N}_B$ .

When the gravitational strength component $\textbf{b}_m$ produced by the revolution of celestial body $m$ can be neglected, the strength gradient force $\textbf{N}_B$ will be symmetrical with respect to the symmetric axis A-A(M) of the gravitational strength component $\textbf{b}_M$ produced by the rotation of celestial body $M$. At the moment the force $\textbf{N}_B$ in the symmetric axis A-A(M) is comparatively strong, and that in the equatorial plane of celestial body $M$ is comparatively weak. When the gravitational strength component $\textbf{b}_M$ produced by the rotation of celestial body $M$ can be neglected, the strength gradient force $\textbf{N}_B$ will be symmetrical with respect to the symmetric axis A-A(m) of the gravitational strength component $\textbf{b}_m$ produced by the revolution of celestial body $m$ . Right now the force $\textbf{N}_B$ in the symmetric axis A-A(m) is comparatively strong, and that in the orbital plane of celestial body $m$ is comparatively weak. When the movements of two celestial bodies, $M$ and $m$ , produce jointly the gravitational strength component $\textbf{b}$ , and there is one small angle between the orbital plane of celestial body $m$ and the equatorial plane of celestial body $M$, therefore the force $\textbf{N}_B$ around the rotational axis of the celestial body $M$ is comparatively strong, and that around the equatorial plane of celestial body $M$ is comparatively weak.

The surface of celestial body $M$ is permeated with some kinds of particles due to the existence of certain production mechanisms. Therefore it enables a fraction of particles to catch a chance to escape temporarily from the gravitational restriction of celestial body $M$. The most of the escaping particles will be dropped back to the surface of celestial body $M$ in a very short time due to the attraction of gravity of celestial body $M$, although a few of these escaping particles may take a much longer time to be dropped back to the surface. As a result, before dropping back to the surface of celestial body $M$, a little of the escaping particles are able to be accelerated by the strength gradient force $\textbf{N}_B$ , to move outward along the direction of $(\nabla b^2 / \mu_g)$ . Under the influence of the force $\textbf{N}_B$ , these escaping particles will be accelerated continually to form the astrophysical jets finally. Meanwhile the strength gradient force $\textbf{N}_B$ is able to thrust any ingredient of matter coming from the celestial body $M$, including the charged and neutral particles etc.

According to the symmetry of strength gradient force $\textbf{N}_B$ , the launching particles will be symmetrical with respect to the precessional axis A-A near the rotational axis of celestial body $M$. In the region around the equatorial plane of celestial body $M$, the force $\textbf{N}_B$ is comparatively weak, therefore the velocity of emitting particle is comparatively slow, and even emerging to pile up the particles to a certain extent. In the region around the rotational axis of the celestial body $M$, the force $\textbf{N}_B$ is comparatively strong, and the velocity of emitting particle is comparatively swift, and even emerging the superfast jet (that is the bipolar jets) to a certain extent. Moreover the revolution of celestial body $m$ has an enhancive influence on the existing gravitational strength component $\textbf{b}$ and the strength gradient force $\textbf{N}_B$ . And the strength gradient force $\textbf{N}_B$ may possess the periodic fluctuating component, due to the revolution period of celestial body $m$ .

In the case of $\mathbb{N}_g = 0$, for the neutral particle, the gravitational acceleration of one test particle corresponds to its linear acceleration for $\textbf{N}_1^i = 0$; According to $\textbf{N}_1 = 0$, the gravitational strength component $\textbf{b}$ , which one test particle undertakes, corresponds to the double of the precessional angular velocity of the particle when $k = 2$. On the basis of the above inferences, for the charged particle, it is necessary to consider the influence of the electromagnetic force and the Larmor precession for one charged particle and so on. In other words, in the gravitational field of the celestial body $M$ with the gravitational strength component $\textbf{b}$ and gravitational acceleration, the test particle must possess the precessional angular velocity and linear acceleration, to meet the requirement of $\textbf{N}_1^i = 0$ and $\textbf{N}_1 = 0$ simultaneously.

The above model is supposed to be appropriate to the jet phenomenon of the pulsar and black hole and so forth. Because of the existence of certain production mechanisms, a fraction of particles are capable of escaping temporarily from partial regions of the surface of pulsar (or black hole and so forth). Furthermore under the influence of the force $\textbf{N}_B$ , a few particles are able to be accelerated continually to outburst finally. Observing far away from the pulsar and black hole and so forth, one may find the accretion disk around the equatorial plane of celestial body, and the astrophysical jets around the rotational axis of celestial body.

In a similar way, the above model is able to be applied partially to explain the phenomenon of solar wind. The sun can be considered as one celestial body $M$ with the rotation. The bursting of solar flare enables some particles to escape from partial regions of the solar surface. Subsequently, under the influence of the force $\textbf{N}_B$ , a few escaping particles will be emitted along the direction of $(\nabla b^2 / \mu_g)$£¬ and accelerated continually. The emitting particles include the charged and neutral particles and so forth. According to the above analysis, the solar wind is symmetrical approximately with respect to the rotational axis and the equatorial plane simultaneously. Around the equatorial plane of the sun, the force $\textbf{N}_B$ exerting on the particle of solar wind is comparatively weak, and the velocity of the particle is comparatively slow. In the region around the rotational axis of the sun, the force $\textbf{N}_B$ exerting on the emitting particle is comparatively strong, and the corresponding velocity of particle is comparatively swift. Meanwhile it is necessary to consider the influence of solar rotation and of solar precession on the formation of the solar wind, and the impact of the plasma feature of solar wind on the solar jets also.

\subsection{\label{sec:level2}Jet source}

In the gravitational and electromagnetic fields of the celestial body $M$ with the rotation and electric charge, there are comparatively strong gravitational strength and electromagnetic strength within the spatial regions near the celestial body $M$. From Eq.(11), the strong gravitational strength and electromagnetic strength will result in the decreasing of the gravitational mass of the test particle. The gravitational mass $m_g$ of the test particle $m$ is approximately to, $m_g = ( m^\prime + m ) + W_E / (k_p v_0^2)$ . Along with the test particle $m$ approaching gradually to the celestial body $M$, the gravitational mass $m_g$ will decrease continuously, until the gravitational mass $m_g$ becomes a negative number. In the case of the gravitational mass $m_g$ is one negative number, the test particle $m$ will experience the repulsive force of the celestial body $M$, and move gradually away from the celestial body $M$, until the gravitational mass $m_g$ becomes a positive number again. And then the test particle $m$ will experience the attractive force of the celestial body $M$ again. The change process may be repeated again and again.

From the expression of the gravitational mass $m_g$ , it is able to estimate the maximum radius of the spherical celestial body $M$, without the rotation and electric charge. The celestial body $M$ is supposed to consist of one uniform particle $m$ with the same density, the gravitational mass of particle on the surface of celestial body $M$ is $m_g$. When the radius of spherical celestial body $M$ becomes more and more enormous, the gravitational mass $m_g$ will be decreased gradually due to the increasing of gravitational acceleration, until to $m_g = 0$. At the moment the radius of celestial body $M$ reaches to its maximum value finally. Obviously the radius of celestial body $M$ has one different maximum value for one different particle $m$ . It means that the pulsar and black hole and so forth are possible to possess their maximum radiuses respectively. When the celestial body $M$ owns the rotation, electric charge, and variable density, the maximum radius of the spherical celestial body $M$ needs to be modified.

In the region near the surface of celestial body $M$ with an intense gravitational strength, there are a few probabilities of generating the particle antiparticle pair. The antiparticle with the negative gravitational mass will be pushed away from the surface of celestial body $M$ directly. In the strong gravitational field, the gravitational mass of particle will be decreased, and the same is true of the gravitational mass of antiparticle. To a certain extent, the gravitational mass of particle is able to be one negative number also, and then to be pushed away from the surface of celestial body $M$ with the strong gravitational field. Therefore the surface and near region of celestial body $M$ are possible to permeate with some kinds of particles escaping temporarily. Certainly there may be other kinds of production mechanisms (similar to the solar flare), enabling the particles to escape temporarily from the surface of celestial body $M$ too. In a word, the astrophysical jet may be come from the surface, the near region, or the inner accretion disk of celestial body $M$.

In the gravitational and electromagnetic fields of the celestial body $M$ with the rotation and electric charge, one test particle $m$ is moving towards the celestial body $M$. Along with the test particle $m$ approaching the celestial body $M$ gradually, the gravitational mass $m_g$ will decrease continuously, until the gravitational mass $m_g$ becomes a negative number, and then the test particle $m$ moves gradually away from the celestial body $M$. At the moment, the minimum distance between the test particle $m$ and the celestial body $M$ is one limit, which the test particle $m$ is incapable of going beyond forever. In the course of moving towards the celestial body $M$, only the test particle $m$ , for which the inertial mass is large enough, is able to reach the surface of celestial body $M$, because the gravitational mass $m_g$ remains in the positive number all the time. Obviously the minimum distance is different for the test particle $m$ with different inertial mass. In other words, within the space between the celestial body $M$ and the inner accretion disk, there is one vacuum region within which the tiny-mass particle cannot exist, for this kind of particle.

\begin{table}[h]
\tbl{Some dependency relationships between the dynamic quantity (the component $\textbf{b}$) and the movement quantity (the astrophysical jet) in the gravitational and electromagnetic fields described with the complex octonion.}
{\begin{tabular}{@{}lll@{}}
\hline
dynamics                     &  component~$\textbf{b}$                                   &  astrophysical~jet                      \\
\hline
launching                    &  force~term,~$\nabla b^2 / \mu_g $                        &  continuously~accelerating              \\

polar~axis                   &  precessional~axis                                        &  near~to~the~rotational~axis            \\

polarity                     &  symmetrical~with~respective~to~the                       &  bipolarity~with~equilong~jets          \\
                             &  ~~equatorial~plane~and~precessional~axis                 &                                         \\

collimation                  &  precessional~angular~velocity                            &  linear~motion~with~the~jet-rotation    \\

intermittency                &  emergence~of~the~linear~momentum                         &  emission~of~the~jet                    \\

\hline
\end{tabular}}
\end{table}

\subsection{\label{sec:level2}Jet property}

According to the feature of strength gradient force and the model of planar circular motion similar to the magnetic dipole moment, it is able to deduce a few fundamental characteristics of the astrophysical jet.

\subsubsection{\label{sec:level3}Jet dynamics}

Within the components of force $\textbf{N}_B$ , the gradient of norm of the gravitational strength component $\textbf{b}$ has a comparatively strong influence on the astrophysical jet, according to Eqs. (12) and (14). When the gradient of norm of the gravitational strength component $\textbf{b}$ is large, the strength gradient force is strong also, and the emitting distance is long. Moreover the collision strength of matters within the astrophysical jet is great, and the luminosity is high. When the gradient of norm of the gravitational strength component $\textbf{b}$ is small, the strength gradient force is weak also, and the emitting distance is short. Moreover the collision strength of matters within the astrophysical jet is inferior, and the luminosity is low.

When the gravitational strength component $\textbf{b}$ is weak, which produced by the rotation of celestial body $M$ and the revolution of particle $m$ within the accretion disk, it may be unable to emit the astrophysical jet. When the gravitational strength component $\textbf{b}$ is strong, which produced by the rotation of celestial body $M$ and the revolution of particle $m$ within the accretion disk, it may be able to emit the astrophysical jet, such as AGN.

The comparatively strong $\nabla b^2$ will be concentrated mainly on the region near the rotational axis of celestial body $M$, enabling the emitting particles to be accelerated continuously in the region. On the precessional axis of celestial body $M$, the term $\textbf{N}_B$ becomes one force to be able to accelerate continuously the particles to move outwards. Theoretically the force can be extended to the infinity along the precessional axis. Moreover it is supposed that there may be the magnetic-like permeability, $\mu_g^\prime$ , and the relative-like permeability, $\mu_{gr}$ , to satisfy the formula, $\mu_g^\prime = \mu_{gr} \mu_g$ . And these coefficients of gravitational field result in the gravitational strength component $\textbf{b}$ inside the celestial body $M$
(and even the astrophysical jets) to be magnified to a certain extent, reaching up to the comparatively high levels in the region around the rotational axis.

In the planar circular motion similar to the magnetic dipole moment, the distribution of gravitational strength component $\textbf{b}$ , is symmetrical with respect to the precessional axis of celestial body $M$. Obviously only the particles emitting along the precessional axis can be accelerated continuously for a very long time, generating the precessional angular velocity for the test particle, and enhancing further the collimation and stability of the astrophysical jet. But along other directions, the emitting particles can only be accelerated for a short time.

\subsubsection{\label{sec:level3}Matter ingredient of jet}

The inferences of Eqs.(12) and (14) reveal that the strength gradient force relates only to the gradient of the norm of gravitational strength and of electromagnetic strength, and is independent of not only the direction of field strength but also the mass and electric charge for the test particle. It means that any kind of particle (of ordinary matter, and even of dark matter), which is able to be sensible of the influence of force, may be one of ingredients of astrophysical jets. The matter of astrophysical jets may come from the surface, or the near region, or the inner accretion disk of celestial body $M$.

\subsubsection{\label{sec:level3}Jet collimation}

According to the gravitational field equations, the rotation of celestial body $M$ and the revolution of orbital particle $m$ may produce jointly the gravitational strength component $\textbf{b}$ with sufficient intensity. In the case of $\mathbb{N}_g = 0$, from $\textbf{N}_1 = 0$, the gravitational strength component $\textbf{b}$ , which one test particle undertakes, corresponds to the double of the precessional angular velocity of the particle when $k = 2$. In other words, the astrophysical jet is emitted along the symmetrical axis (or precessional axis) of gravitational strength component $\textbf{b}$ only, rather than along the rotational axis of celestial body $M$. Of course, the angle between the precessional axis and the rotational axis may be tiny.

Obviously the inference of $\textbf{N}_1 = 0$ reveals that the emitting jet possesses one precessional angular velocity. It states that the astrophysical jet owns not only the linear acceleration but also the precessional angular velocity. Even if neglecting the contribution of the rotational velocity of celestial body $M$, the astrophysical jet with the precessional angular velocity is still able to maintain its stability and collimation, emitting to somewhere very far away (Table III).

In the region around the rotational axis of celestial body $M$, the precessional angular velocity is comparatively quick. Meanwhile in the region around the equatorial plane of celestial body $M$, the precessional angular velocity is comparatively slow. As a result they form the differential of precessional angular velocity, and then may make partially a contribution to generating differential rotations among some regions of celestial body $M$.

\subsubsection{\label{sec:level3}Jet polarity}

The distribution of gravitational strength component $\textbf{b}$ will determine the astrophysical jet to be unipolar or bipolar. In the vacuum, there is one celestial body $M$ with the rotational velocity. When the distribution of gravitational strength component $\textbf{b}$ produced by the celestial body $M$ is symmetrical with respect to not only the symmetric axis A-A but also the equatorial plane, the astrophysical jets will be bipolar and equilong. When the distribution of gravitational strength component $\textbf{b}$ is unsymmetrical with respect to the equatorial plane, the astrophysical jets are still bipolar, but one jet may be longer than the other. When the distribution of gravitational strength component $\textbf{b}$ departs rigorously from the symmetry with respect to the equatorial plane, the astrophysical jets may be unipolar, because one of jets may be too short to be found. Moreover the revolution of celestial body $m$, orbiting around the celestial body $M$, will be capable of impacting the length of jet also.

In the case of the distribution of gravitational strength component $\textbf{b}$ being symmetrical with respect to not only the symmetric axis A-A but also the equatorial plane, the rotational velocity of celestial body $M$ will be able to impact the length of astrophysical jet. According to $\textbf{N}_1 = 0$, the emitting direction of one jet is parallel to the direction of precessional angular velocity, while the emitting direction of the other jet is antiparallel to the direction of precessional angular velocity. It means that the rotational speed $\omega_0$ of celestial body $M$ will speed up the precessional speed of one jet partially, while slow down that of the other partially. As a result, the collimation and stability, which related with the precessional speed, of two jets appear to be diversified, and then the emitting distances are different correspondingly. In other words, two jets may be equilong, when $\omega_0 = 0$. When $\omega_0$ is comparatively large, one jet may be longer than the other. When $\omega_0$ is awfully large, it is possible that only one of two jets could be observed. In a similar way, when the distribution of gravitational strength component $\textbf{b}$ is unsymmetrical with respect to the equatorial plane, the rotational velocity of celestial body $M$ will be able to impact the length of astrophysical jet. Furthermore the orbital speed of celestial body $m$ (or accretion disk) has an influence on the length of jet too.

Moreover the revolution of celestial body $m$ (or accretion disk) around the celestial body $M$ may produce the gyroscopic torque too, resulting in other kinds of precessional motions of astrophysical jets. Meanwhile the astrophysical jet can be considered as the high-temperature plasma, at the moment it is necessary to consider the pinch effect of plasma within the magnetic field produced by the launching of astrophysical jet. In a word, the precessional motion of astrophysical jet is diversiform, enabling the jet to appear to be multiform.

\subsubsection{\label{sec:level3}Jet intermittency}

The astrophysical jet is associated with the orbital motion of the celestial body $m$ around the celestial body $M$. The gravitational strength has an influence on the variation of gravitational mass. The gravitational acceleration $\textbf{g}$ will decrease the gravitational mass of test particle, while the gravitational strength component $\textbf{b}$ will increase that. Therefore the discrete and comparatively heavy celestial body is able to be fallen into the celestial body $M$ surrounded by the accretion disk, but the tiny-mass particle is uneasy to be fallen into that. It means that the astrophysical jet is intermittent, when the comparatively heavy celestial body is fallen into the celestial body $M$. When a series of celestial bodies with comparatively small mass are fallen into the celestial body $M$, the astrophysical jet is able to be continuously.

Because the gravitational strength is capable of varying the gravitational mass of particle, the smaller the celestial body is, the more difficult it is close to the celestial body $M$ surrounded by the accretion disk. It leads to produce the clear-cut boundary between the inner accretion disk with the celestial body $M$. During one heavy celestial body $m$ is fallen into the celestial body $M$ gradually, it may induce one strong component $\textbf{b}$ . And it results in the gravitational mass of celestial body with comparatively small mass, within the near region around the celestial body $M$, is restored and even increased, therefore one part of celestial bodies with comparatively small mass will be able to be fallen to the celestial body $M$, while other part of them may be ejected outwards.

The astrophysical jet may be interim, and even it is possible to appear to the long-playing intermission. Since the heavy celestial body fallen into the celestial body $M$ is discrete, and its orbit is periodic. And it enables the emergence of the maximum value of component $\textbf{b}$ may be periodic (or intermittent), and then the astrophysical jet is periodic (or intermittent) accordingly. Furthermore the astrophysical jet emits along the precessional axis, resulting in the pattern of astrophysical jet to be in an unceasing precessional motion all along.

\section{\label{sec:level1}Experiment proposal}

In the laboratory experiment, it is very difficult to validate the influence of the gravitational strength (the $\textbf{g}$ and $\textbf{b}$ ) on the gravitational mass $m_g$ and strength gradient force $\textbf{N}_B$ in the complex octonion space. Only the influence may be observed in the astrophysical phenomenon. Fortunately it is able to validate directly the influence of the electromagnetic strength (the $\textbf{E}$ and $\textbf{B}$ ) on the gravitational mass $m_g$ and its strength gradient force $\textbf{N}_B$ in the laboratory. For instance, in the strong (uniform or non-uniform) magnetic field, it is capable of measuring the variation $m^\prime$ of gravitational mass and its influence on the strength gradient force $\textbf{N}_B$ . By means of the observation of the E\"{o}tv\"{o}s experiment, optical tweezers, magnetic mirror and so forth in the strong magnetic field, it is able to determine the influencing degree of the strength gradient force $\textbf{N}_B$ on the neutral particle and relevant motions.

(1) E\"{o}tv\"{o}s experiment in the strong magnetic field. According to the force equilibrium equation, $\textbf{N}_1^i = 0$, the gravitational mass is approximately written as, $m_g = ( m^\prime + m  ) + W_E / (k_p v_0^2)$ . However the E\"{o}tv\"{o}s experiment has never been inspected under the electromagnetic environment till to now. Consequently it is necessary to validate the E\"{o}tv\"{o}s experiment in the strong magnetic field. When the distribution of strong magnetic field is uniform, the variation of magnetic flux density will alter the gravitational mass. In case the distribution of strong magnetic field is non-uniform, the variation of magnetic flux density will result in not only the alteration of gravitational mass but also the emergence of strength gradient force $\textbf{N}_B$ . As a result, the strong magnetic field must break the existing state of force equilibrium, transferring the existing equilibrium position of the neutral particle. Furthermore, on the basis of existing E\"{o}tv\"{o}s experiments, it is feasible to validate the E\"{o}tv\"{o}s experiment via applying strong magnetic fields in the experimental technique.

(2) Optical tweezers in the strong magnetic field. In the optical tweezers, it is able to apply the tiny gradient force of optical tweezers to manipulate the biologic cell and so forth. According to the feature of strength gradient force $\textbf{N}_B$ , the force $\textbf{N}_B$ is capable of generating an arbitrary angle with the gradient force of optical tweezers. And that the $\textbf{N}_B$ may be either larger or smaller than the gradient force of optical tweezers. Differing from the optical tweezers, it is unnecessary to give out light to produce the force $\textbf{N}_B$ , enabling the experimental observation to be much distinct. On the basis of existing experiments of optical tweezers, it is able to measure the extra motion and force of the neutral particle in the optical tweezers, due to the application of the strong non-uniform magnetic field. Furthermore it may be able to contrast related variations, before and after applying the strong non-uniform magnetic field.

(3) Magnetic mirror in the strong magnetic field. In the magnetic mirror, the charged particle comes and goes between two magnetic loops. However the neutral particle is also able to come and go between two sides of one single magnetic loop in the magnetic mirror, according to the above strength gradient force $\textbf{N}_B$ . Building on past achievements of magnetic mirror experiment, generating the strong non-uniform magnetic field by one single magnetic loop, it is capable of observing the come-and-go motion of the neutral particle in the magnetic mirror.

The validation of above experiment proposal will be of benefit to investigate the further features of strength gradient force and of astrophysical jets.

\section{\label{sec:level1}Conclusions and Discussions}

In the paper, the complex quaternion space for the electromagnetic field is independent of that for the gravitational field. These two complex quaternion spaces can compose one complex octonion space. In the complex octonion space, it is able to define the octonion field potential, octonion field strength, and octonion field source related with the electromagnetic and gravitational fields, and then deduce the angular momentum, torque, force and so forth. The force includes the strength gradient force and so on.

Either the electromagnetic strength or the gravitational strength makes a direct contribution to the strength gradient force. In the planar circular motion similar to the magnetic dipole moment, when there only is the gravitational strength component $\textbf{b}$ , the strength gradient force exerts the thrust on the astrophysical jet along the precessional axis outwards. Moreover the force $\textbf{N}_B$ is independent of the mass and electric charge, and then is capable of ejecting continuously any particle of ordinary matter to the infinity along the precessional axis theoretically.

According to the feature of strength gradient force, it is able to infer a few characteristics of the astrophysical jet as follows. (1) When there is the gravitational strength component $\textbf{b}$ only, the force $\textbf{N}_B$ presses on the astrophysical jet to launch outwards. (2) All of astrophysical jets are bipolar, due to the closed force line of the component $\textbf{b}$ , which is similar to that of the magnetic flux density $\textbf{B}$ . (3) Two oppositely directed jets of the astrophysical jet may be either equilong or non-equilong, due to different symmetries of $(\nabla b^2 / \mu_g)$; The length of astrophysical jet may be either long or short, due to different norms of the component $\textbf{b}$ . (4) The existence of the component $\textbf{b}$ enables the astrophysical jet to own the precessional angular velocity, and then to maintain the collimation and stability of jets. (5) The time interval of the emergence of linear momentum determines the period of time of the production of component $\textbf{b}$ , resulting in the astrophysical jet to be occurred with one time limit.

Experimentally it is very tough to inspect directly the strength gradient force produced by the gravitational strength. However it is comparatively easy to verify directly the strength gradient force produced by the electromagnetic strength. When there is the magnetic flux density $\textbf{B}$ only, one can check out the influence of strength gradient force on the E\"{o}tv\"{o}s experiment, optical tweezers, magnetic mirror and so forth. On the basis of above existing experiments, it is feasible to achieve these experiments technically.

It should be noted that the paper discussed only some simple cases, of which the strength gradient force is chosen as the dynamic of astrophysical jet, and a few relevant inferences and so forth. However it clearly states that the feature of strength gradient force is able to explain availably a majority of movement phenomena of astrophysical jet, including the oppositely directed jets, matter ingredient of jet, launching, collimation, stability, emitting along the precessional axis, continuously accelerating and so on. In the following researches, we plan to explore further the dynamic feature of astrophysical jet via the strength gradient force theoretically, and prove the influence of the strength gradient force produced by the magnetic flux density on the existing experiments, and apply the strength gradient force to develop the decelerator device for the particle trapping.

\section*{Acknowledgments}
This project was supported partially by the National Natural Science Foundation of China under grant number 60677039.



\begin{thebibliography}{0}    



\bibitem{majernik}
      V. Majernik,
      {\it Adv. Appl. Clifford Al.\/}
      {\bf 9} (1999) 119.

\bibitem{honig}
      W. M. Honig,
      {\it Lett. Nuovo Cimento\/}
      {\bf 19} (1977) 137.

\bibitem{singh}
      A. Singh,
      {\it Lett. Nuovo Cimento\/}
      {\bf 31} (1981) 145.

\bibitem{edmonds}
      J. D. Edmonds,
      {\it Int. J. Theor. Phys.\/}
      {\bf 10} (1974) 115.

\bibitem{morita}
      K. Morita,
      {\it Prog. Theor. Phys.\/}
      {\bf 117} (2007) 501.

\bibitem{grusky}
      S. M. Grusky, K. V. Khmelnytskaya, and V. V. Kravchenko,
      {\it J. Phys. A\/}
      {\bf 37} (2004) 4641.

\bibitem{anastassiu}
      H. T. Anastassiu, P. E. Atlamazoglou, and D. I. Kaklamani,
      {\it IEEE T. Antenn. Propag.\/}
      {\bf 51} (2003) 2130.

\bibitem{winans}
      J. G. Winans,
      {\it Found. Phys.\/}
      {\bf 7} (1977) 341.

\bibitem{demir}
      S. Demir, M. Tanisli, and T. Tolan,
      {\it Int. J. Mod. Phys. A,\/}
      {\bf 28} (2013) 1350112.

\bibitem{doria}
      F. A. Doria,
      {\it Lett. Nuovo Cimento\/}
      {\bf 14} (1975) 480.

\bibitem{rawat}
      A. S. Rawat and O. P. S. Negi,
      {\it Int. J. Theor. Phys.\/}
      {\bf 51} (2012) 738.

\bibitem{gogberashvili}
      M. Gogberashvili,
      {\it J. Phys. A\/}
      {\bf 39} (2006) 7099.

\bibitem{mironov}
      V. L. Mironov and S. V. Mironov,
      {\it J. Math. Phys.\/}
      {\bf 50} (2009) 012901.

\bibitem{dehnen}
      O. P. S. Negi, H. Dehnen, G. Karnatak, and P. S. Bisht,
      {\it Int. J. Theor. Phys.\/}
      {\bf 50} (2011) 1908.

\bibitem{weng1}
      Z.-H. Weng,
      {\it Adv. Math. Phys.\/}
      {\bf 2014} (2014) 450262.

\bibitem{weng2}
      Z.-H. Weng,
      {\it AIP Adv.\/}
      {\bf 4} (2014) 087103.

\bibitem{yu}
      L.-Y. Yu and Y.-L. Sheng,
      {\it Opt. Express\/}
      {\bf 22} (2014) 7953.

\bibitem{bagryansky}
      P. A. Bagryansky, Y. V. Kovalenko, V. Y. Savkin, A. L. Solomakhin, and D. V. Yakovlev,
      {\it Nucl. Fusion\/}
      {\bf 54} (2014) 082001.

\bibitem{cao}
      X. Cao and D. R. Jiang,
      {\it Mon. Not. R. Astron. Soc.\/}
      {\bf 320} (2001) 347.

\bibitem{cherepashchuk}
      A. M. Cherepashchuk, R. A. Sunyaev, S. V. Molkov, E. A. Antokhina, K. A. Postnov, and A. I. Bogomazov,
      {\it Mon. Not. R. Astron. Soc.\/}
      {\bf 436} (2013) 2004.

\bibitem{alves}
      E. P. Alves, T. Grismayer, R. A. Fonseca, and L. O. Silva,
      {\it New J. Phys.\/}
      {\bf 16} (2014) 035007.




\end{thebibliography}
\end{document}